\documentclass[12pt]{article}
\usepackage{eurosym}
\usepackage{graphicx}
\usepackage[utf8]{inputenc}
\usepackage[T1]{fontenc}
\usepackage{indentfirst}
\usepackage[margin=0.6in,nomarginpar]{geometry}
\usepackage[final]{hyperref}
\usepackage{amsmath}
\usepackage{hyperref}
\usepackage{cite}
\usepackage{subcaption}
\usepackage{caption}
\usepackage{amssymb}
\usepackage{multirow}
\usepackage[table]{xcolor}
\usepackage{orcidlink}

\hypersetup{
colorlinks=true,
linkcolor=blue,
citecolor=blue,
filecolor=magenta,
urlcolor=blue
}

\begin{document}

\title{Time-dependent Dunkl-Pauli Oscillator}
\author{A. Benchikha \orcidlink{0009-0003-9848-0822}\thanks{%
benchikha4@yahoo.fr} \\
$^{1}$D\'{e}partement de EC, Facult\'{e} de SNV, Universit\'{e} Constantine 1 Fr\`{e}res
Mentouri,\\
Constantine, Algeria.\\
$^{2}$Laboratoire de Physique Math\'{e}matique et Subatomique,\\
LPMS, Universit\'{e} Constantine 1 Fr\`{e}res  Mentouri, Constantine,
Algeria. \and  B. Hamil \orcidlink{0000-0002-7043-6104} \thanks{%
hamilbilel@gmail.com} \\
Laboratoire de Physique Math\'{e}matique et Subatomique,\\
LPMS, Universit\'{e} Constantine 1 Fr\`{e}res Mentouri, Constantine,
Algeria.  \and B. C. L\"{u}tf\"{u}o\u{g}lu \orcidlink{0000-0001-6467-5005}\thanks{%
bekir.lutfuoglu@uhk.cz (Corresponding author)}, \\
Department of Physics, Faculty of Science, University of Hradec Kralove,\\
Rokitanskeho 62/26, Hradec Kralove, 500 03, Czech Republic.}
\date{\today }
\maketitle

\begin{abstract}
This study explores the time-dependent Dunkl-Pauli oscillator in two dimensions. We constructed the Dunkl-Pauli Hamiltonian, which incorporates a time-varying magnetic field and a harmonic oscillator characterized by time-dependent mass and frequency, initially in Cartesian coordinates. Subsequently, we reformulated the Hamiltonian in polar coordinates and analyzed the eigenvalues and eigenfunctions of the Dunkl angular operator, deriving exact solutions using the Lewis-Riesenfeld invariant method. Our findings regarding the total quantum phase factor and wave functions reveal the significant impact of Dunkl operators on quantum systems, providing precise expressions for wave functions and energy eigenvalues. This work enhances the understanding of quantum systems with deformed symmetries and suggests avenues for future research in quantum mechanics and mathematical physics.
\end{abstract}



\section{Introduction}
The Pauli equation, developed by Wolfgang Pauli as an extension of the Schr\"{o}dinger equation, integrates electron spin and magnetic field interactions to describe the dynamics of non-relativistic spin-1/2 particles, such as electrons, in electromagnetic fields \cite{1,2,3,4,5,6,7,8,9}. This equation has had a profound impact on quantum chemistry and particle physics, aiding in the classification of Lie symmetries for charged particles and quasi-relativistic Schr\"{o}dinger equations \cite{10,11,12,13}. It has also facilitated the discovery of new superintegrable systems exhibiting spin-orbit coupling, laying the theoretical foundation for atomic structure and the periodic table through quantum spin theory and the Pauli exclusion principle \cite{13,14,15,16,17,18,19,20,22}. Additionally, it supports Bohr's model of the hydrogen atom, which was previously regarded as empirical \cite{23,24}. Accurate solutions to the Pauli equation are crucial for predictive quantum chemistry, enabling precise predictions of multi-electron atoms and molecules \cite{25, 26, 27, 28}.

Dunkl operators, defined as commuting differential-difference operators linked to finite reflection groups \cite{Dunkl1989}, serve as a generalization of partial derivatives, enabling a broader framework for analyzing functions and their symmetries in pure mathematics \cite{Dunkl2008, Dunkl2014} and theoretical physics \cite{Kakei1996, Lapointe1996, Hikami1996}. Notably, the Dunkl-like operators first appeared in physics in 1951 through Yang's seminal work \cite{Yang1951}, drawing inspiration from the pioneering contributions of Wigner from the previous year \cite{Wigner1950}. Yang's work opened new frameworks in quantum field theory by proposing a generalized field quantization \cite{Green1953}, which significantly aided in the introduction of color degrees of freedom and the development of quantum chromodynamics \cite{Greenberg1964}. Subsequently, reflection operator deformed algebras have been employed in various subareas of physics, including particle physics and statistical mechanics, further enhancing our understanding of symmetries and interactions in these fields \cite{Plyushchay1994, Plyushchay1996, Plyushchay1997, Gamboa1999,   Plyushchay2000, Klishevich2001, Horvathy2004, Rodrigues2009, Horvathy2010, Ubriaco2014, Luo2020}.

In the past decade, Dunkl operators have increasingly been employed as substitutes for ordinary partial derivatives, leading to a richer algebraic structure in various quantum systems, both relativistic and nonrelativistic. This substitution not only enhances the understanding of solvability and integrability but also introduces new symmetries, providing deeper insights into the spectral properties and dynamics of these systems \cite{Genest20131, Genest20132, Genest20133, Genest20141, Genest20142, Vincent3, Isaac2016, Salazar2017, Salazar2018, Mota2019, Ghazouani2019, Ghazouani2020, Ghazouani2021, Hamil20221, Hamil20222, Hamil20223, Samira, 43, 46, 50, 51, Junker2023, Quesne2024, 66, 67, 67a, Schulze20243, Schulze20244, Schulze20245, Mota20241, Mota20242, Benzair20241, Benzair20242, Junker2024}. Recently, two authors of the present manuscript have investigated the Pauli equation within the Dunkl formalism in the presence of a magnetic field, deriving the Dunkl-Pauli Hamiltonian, which incorporates terms involving the Dunkl-Laplacian and additional potentials for magnetic field and spin interactions, with the solutions successfully presented in \cite{Bouguerne20241}. However, that was the stationary scenario, and in the non-stationary case, the dynamics are expected to reveal further complexities, which remain an open area of investigation. To address time-dependent scenarios, one can extend our previous analysis by considering a magnetic field that varies with time, along with a harmonic oscillator featuring both time-dependent mass and frequency potential. Such an extension results in a modified Hamiltonian that encapsulates the dynamic interactions between these variables. From there, one can express the wave function in polar coordinates and introduce the Dunkl-angular operator to facilitate solving the Hamiltonian. The eigenfunctions and eigenvalues of the Dunkl-angular operator are expected to play a central role in constructing solutions to the Dunkl-Pauli equation. Consequently, one can investigate cases with different eigenvalues of the reflection operators and derive the corresponding radial wave functions. Additionally, one can employ the invariant method to identify the time-dependent invariant and use unitary transformations to simplify the eigenvalue problem \cite{68, 69, 70, 71, 72, 73, 74, 75}.

The Pauli equation and the Pauli oscillator are both concepts in quantum mechanics that describe different behaviors of a spin-1/2 particle. The Pauli equation encompasses the general motion of the particle in electromagnetic fields. In contrast, the Pauli oscillator specifically pertains to the dynamics of such particles within a harmonic oscillator, incorporating spin and a magnetic field in a quadratic potential. Notably, the solution of the Pauli oscillator in conjunction with the Snyder-de Sitter algebra and noncommutative space yields intriguing predictions that further our understanding of quantum systems \cite{Santos2011, Heddar2021, Guerira2022}. Motivated by the facts summarized above, in this study, we aim to explain the dynamics of quantum systems described by the time-dependent Pauli oscillator within the framework of Dunkl formalism. We believe that the solution will provide new insights into the interplay between symmetry, algebra, and quantum dynamics, paving the way for potential applications in studying complex quantum systems with enhanced symmetry properties.

The paper is organized as follows: In Sec. \ref{sec2}, we derive the two-dimensional time-dependent Dunkl-Pauli Hamiltonian, first in Cartesian coordinates and then in polar coordinates. In Sec. \ref{sec3}, we employ the Lewis-Riesenfeld invariant method to investigate the wave function solutions. Subsequently, in Sec. \ref{sec4}, we discuss the total quantum phase factor and present the exact form of the wave functions. Finally, we conclude the manuscript with a brief conclusion section.

\section{Time-dependent Dunkl-Pauli oscillator in two dimensions} \label{sec2}

We consider the time-dependent Pauli equation with a harmonic oscillator potential in two dimensions, which represents the non-relativistic limit of the Dirac equation in standard quantum mechanics:
\begin{equation}
\left[ \frac{1}{2m\left( t\right) }\left( \overrightarrow{\sigma _{j}} \cdot \overrightarrow{\pi _{j}}\right) ^{2}+\frac{1}{2}m\left( t\right) \omega \left( t\right) r^{2}\right] \psi \left( \overrightarrow{r},t\right) =i\frac{\partial }{\partial t}\psi \left( \overrightarrow{r},t\right) ,  \label{1}
\end{equation}
where $m\left( t\right) $ denotes the time-dependent mass, while $\omega \left(t\right) $ corresponds to the time-dependent frequency of the harmonic oscillator. Here, $\sigma _{j}$ refers to the Pauli matrices and $\overrightarrow{\pi }$ are the kinetic momentum terms given by: 
\begin{equation}
\overrightarrow{\pi }=\overrightarrow{p}-\frac{e}{c}\overrightarrow{A}=\left( p_{1}-\frac{e}{c}A_{1},\text{ }p_{2}-\frac{e}{c}A_{2}\right) , \label{2}
\end{equation}%
with charge $e$ and speed of light $c$. In this case, the spinor wave function has two components:
\begin{equation}
\psi =\binom{\psi _{1}}{\psi _{2}}.  \label{3}
\end{equation}
We then adopt the Dunkl formalism by replacing the ordinary momentum operator with the Dunkl momentum operator ($\hbar=1$):
\begin{equation}
p_{j}=\frac{1}{i}D_{j},  \label{4}
\end{equation}%
where the Dunkl derivatives are defined as:%
\begin{equation}
D_{j}=\frac{\partial }{\partial x_{j}}+\frac{\nu _{j}}{x_{j}}\left(1-R_{j}\right) , \qquad {\text{for}} \qquad  j=1,2.   \label{5}
\end{equation}%
In this context, $\nu_j$ refers to the Wigner deformation parameters, while $R_j$ represents the reflection operators. To ensure consistency, the Wigner parameters must satisfy $\nu _{j} > -\frac{1}{2}$, and the reflection operators exhibit the following properties:
\begin{equation}
R_{j}f\left( x_{j}\right) =f\left( -x_{j}\right), \qquad R_{j}x_{i}=-\delta _{ij}x_{j}R_{j}, \qquad \text{and} \qquad R_{j}R_{j}=R_{j}R_{j} .   \label{7}
\end{equation}%
Employing the Dunkl second derivative 
\begin{equation}
D_{j}^{2}=\frac{\partial ^{2}}{\partial x_{j}^{2}}+\frac{2\nu _{j}}{x_{j}}\frac{\partial }{\partial x_{j}}-\frac{\nu _{j}}{x_{j}^{2}}\left(1-R_{j}\right) ,  \label{6}
\end{equation}%
we express the time-dependent Dunkl-Pauli Hamiltonian in Cartesian coordinates as follows:
\begin{equation}
H=\frac{1}{2m\left( t\right) }\bigg[ \pi _{1}^{2}+\pi _{2}^{2}+\sigma _{3}\Big[ \pi _{1},\pi _{2}\Big] \bigg] +\frac{1}{2}m\left( t\right) \omega \left( t\right) r^{2}.  \label{9}
\end{equation}%
We then choose the vector potential in the symmetric gauge, given by:
\begin{equation}
e\overrightarrow{A}=\frac{eB\left( t\right) }{2c}\left( -y\hat{i}+x\hat{j}\right) ,  \label{10}
\end{equation}%
where $\left(\hat{i},\hat{j}\right) $ represents the Cartesian orthonormal basis. By applying the Dunkl-deformed algebra and the definitions provided above, we derive the time-dependent Dunkl-Pauli Hamiltonian in the following form \cite{Bouguerne20241}:
\begin{equation}
H=-\frac{1}{2m\left( t\right) }\triangle _{D}+\frac{m\left( t\right) \Omega^{2}\left( t\right) }{2}\left( x^{2}+y^{2}\right) +i\frac{\omega _{c}\left(t\right) }{2}\left( xD_{2}-yD_{1}\right) -\frac{e}{2m\left( t\right) c}\left( 1+\nu _{1}R_{1}+\nu _{2}R_{2}\right) \sigma _{z}B\left( t\right), \label{11}
\end{equation}%
where the Dunkl-Laplacian is given by: 
\begin{equation}
\triangle _{D}=\frac{\partial ^{2}}{\partial x^{2}}+\frac{\partial ^{2}}{\partial y^{2}}+\frac{2\nu _{1}}{x}\frac{\partial }{\partial x}+\frac{2\nu_{2}}{y}\frac{\partial }{\partial y}-\frac{\nu _{1}}{x^{2}}\left(1-R_{1}\right) -\frac{\nu _{2}}{y^{2}}\left( 1-R_{2}\right).  \label{12}
\end{equation}%
Here, $\omega _{c}\left( t\right) =\frac{eB\left( t\right) }{m(t)c}$ is the time-dependent cyclotron frequency.

\subsection{Solutions in polar coordinates}
At this stage, we prefer to adopt polar coordinates to take advantage of the symmetries, as is customary. Defining the Cartesian coordinates in terms of polar coordinates as 
\begin{equation}
x=r\cos \theta \qquad \text{and} \qquad y=r\sin \theta ,  \label{13}
\end{equation}%
the time-dependent Dunkl-Pauli Hamiltonian takes the following form: 
\begin{align}
H\left( t\right) & =-\frac{1}{2m\left( t\right) }\frac{\partial ^{2}}{\partial r^{2}}-\frac{1+2\nu _{1}+2\nu _{2}}{2m\left( t\right) r}\frac{\partial }{\partial r}+\frac{m\left( t\right) }{2}\Omega ^{2}\left( t\right)r^{2}  \notag \\
& +\frac{\mathcal{J}_{\theta }^{2}-2\nu _{1}\nu _{2}\left(1-R_{1}R_{2}\right) }{2m\left( t\right) r^{2}}+\frac{\omega _{c}\left(t\right) }{2}\Big[ \mathcal{J}_{\theta }-g_{s}\left( 1+\nu _{1}R_{1}+\nu_{2}R_{2}\right) .\mathbf{S}_{z}\Big] ,  \label{14}
\end{align}%
where 
\begin{equation}
\Omega ^{2}\left( t\right) =\omega \left( t\right) ^{2}+\frac{\omega_{c}\left( t\right) ^{2}}{4},  \label{15}
\end{equation}
is the modified harmonic oscillator frequency. Here, $\mathbf{S}_{z}=\frac{\sigma _{z}}{2}$ represents the spin-$\frac{1}{2}$ operator, and $g_{s}$ is the free electron $g$-factor, with $g_{s}=2.0023 $. The Dunkl angular operator, denoted by $\mathcal{J}_{\theta }$, is defined as follows \cite{Genest20131, Genest20132, Genest20141, Genest20142}: 
\begin{equation}
\mathcal{J}_{\theta }=i\left( \frac{\partial }{\partial \theta }+\nu_{2}\cot \theta \left( 1-R_{2}\right) -\nu _{1}\tan \theta \left(1-R_{1}\right) \right) .  \label{17}
\end{equation}%
We then assume the time-dependent wave function takes the form
\begin{equation}
\psi _{m_{s}}\left( r,\theta ,t\right) =\phi \left( r,\theta ,t\right) \chi_{m_{s}},  \label{18}
\end{equation}%
where $\chi _{m_{s}}$ represents the spin function,%
\begin{equation}
\mathbf{S}_{z}\chi _{ms}=\frac{m_{s}}{2}\chi _{m_{s}} \qquad \text{for} \qquad
m_{s}=\pm 1,   \label{19a}
\end{equation}
and 
\begin{equation}
   \chi _{+1}=\binom{1}{0}, \qquad \chi _{-1}=%
\binom{0}{1}.  \label{19b}
\end{equation}
Subsequently, the time-dependent Dunkl-Pauli oscillator equation is expressed as:
\begin{align}
& \left[ -\frac{1}{2m\left( t\right) }\frac{\partial ^{2}}{\partial r^{2}}-\frac{1+2\nu _{1}+2\nu _{2}}{2m\left( t\right) r}\frac{\partial }{\partial r}+\frac{m\left( t\right) }{2}\Omega ^{2}\left( t\right) r^{2}+\frac{\mathcal{J}_{\theta }^{2}-2\nu _{1}\nu _{2}\left( 1-R_{1}R_{2}\right) }{2m\left(t\right) r^{2}}\right.  \notag \\
& \left. +\frac{\omega _{c}\left( t\right) }{2}\Big( \mathcal{J}_{\theta}-m_{s}\left( 1+\nu _{1}R_{1}+\nu _{2}R_{2}\right) \Big) \right] \phi \left( r,\theta ,t\right) =i\frac{\partial }{\partial t}\phi \left( r, \theta, t\right) .  \label{108}
\end{align}%
Here, one can perform the following transformation
\begin{equation}
\phi \left( r,\theta ,t\right) =e^{-\frac{i}{2}\left( \mathcal{J}_{\theta}-m_{s}\left( 1+\nu _{1}R_{1}+\nu _{2}R_{2}\right) \right) \int\limits^{t}\omega _{c}\left( t^{\prime }\right) dt^{\prime }}\mathcal{\digamma }(r,\theta ,t),  \label{109}
\end{equation}
to eliminate the last time-dependent term on the left-hand side of Eq. (\ref{108}), given by  $\frac{\omega_{c}\left( t\right) }{2}\Big( \mathcal{J}_{\theta }-m_{s}\left(1+\nu_{1}R_{1}+\nu _{2}R_{2}\right) \Big)$. Using this transformation, Eq. (\ref{108}) becomes:
\begin{equation}
\left\{ \frac{\mathbf{P}^{2}}{2m\left( t\right) }+\frac{\delta \left( \delta -1\right) -2\nu _{1}\nu _{2}\left( 1-R_{1}R_{2}\right) }{2m\left( t\right)r^{2}}+\frac{m\left( t\right) }{2}\Omega ^{2}\left( t\right) r^{2}\right\}\mathcal{\digamma }(r,\theta ,t)=i\frac{\partial }{\partial t}\mathcal{\digamma }(r,\theta ,t),  \label{a11}
\end{equation}%
where
\begin{eqnarray}
\mathbf{P}^{2} &=&\mathcal{P}^{2}+\frac{\mathcal{J}_{\theta }^{2}}{r^{2}}, \label{111} \\
\mathcal{P} &=&\frac{\hbar }{i}\left[ \frac{\partial }{\partial r}+\frac{\delta }{r}\right] ,  \label{112} \\
\delta &=&\nu _{1}+\nu _{2}+\frac{1}{2},  \label{113}
\end{eqnarray}%
with $\left[ r,\mathcal{P}\right] =i\hbar .$ We observe that the time-dependent Dunkl-Pauli equation mimics the time-dependent
Dunkl-Shr\"odinger equation, as outlined in equation (9) in   \cite{67}.

\section{Lewis-Riesenfeld invariant approach} \label{sec3}

To obtain an exact solution, we utilize the Lewis-Riesenfeld invariant approach  \cite{Riesenfeld1969}, requiring that the Hamiltonian of the system and the invariant $I(t)$ satisfy the Lewis-Riesenfeld invariant equation:
\begin{equation}
\frac{dI(t)}{dt}=\frac{\partial I(t)}{\partial t}+\frac{1}{i}[I(t),\tilde{H}(t)]=0,  \label{114}
\end{equation}%
where the Hamiltonian is given by on the left-side of Eq. \eqref{a11}
\begin{equation}
\tilde{H}(t)=\frac{\mathbf{P}^{2}}{2m\left( t\right) }+\frac{\delta \left(\delta -1\right) -2\nu _{1}\nu _{2}\left( 1-R_{1}R_{2}\right) }{2m\left(t\right) r^{2}}+\frac{m\left( t\right) }{2}\Omega ^{2}\left( t\right) r^{2}.
\label{new Hamiltonian}
\end{equation}
According to this approach the solution of Eq. (\ref{a11}) can be related to the solution of the eigenvalue problem of the invariant
\begin{equation}
I(t)\mathcal{F}(r,\theta ,t)=\varepsilon \mathcal{F}(r,\theta ,t),
\label{I1}
\end{equation}%
with the associated phase factor given by 
\begin{equation}
\mathcal{\digamma }(r,\theta ,t)=e^{i\eta (t)}\mathcal{F}(r,\theta ,t),
\label{115}
\end{equation}%
where the phase $\eta (t)$ can be determined from the equation 
\begin{equation}
\frac{d}{dt}\eta (t)=\Big{\langle} \mathcal{F}(r,\theta ,t)\Big{|}i\frac{\partial }{\partial t}-\tilde{H}\Big{|}\mathcal{F}(r,\theta ,t)\Big{\rangle} .  \label{PH}
\end{equation}%
To find the exact invariants for the system, we define three generators $T_{1}$, $T_{2}$ and $T_{3}$ from the Lie algebra of the group SL(2,R), which satisfy the following commutation relations: 
\begin{equation}
\lbrack T_{1},T_{2}]=-2i\hbar T_{3},\qquad \lbrack T_{2},T_{3}]= 4i\hbar T_{2},\qquad \lbrack T_{1},T_{3}]= - 4i\hbar T_{1}.  \label{37a}
\end{equation}%
and we form the invariant as follows: 
\begin{equation}
I(t)=\frac{1}{2}\Big(\alpha T_{1}+\beta T_{2}+\gamma T_{3}\Big).  \label{37b}
\end{equation}%
Subsequently, by applying Eq. (\ref{114}), we obtain a set of coupled equations whose solutions are expressed in terms of real functions of time, $\alpha$, $\beta$, and $\gamma $:
\begin{equation}
\left\{ \begin{aligned} \alpha &= \rho^{2}, \\ \beta &= \frac{1}{\rho^{2}} + m^{2}\dot{\rho}^{2}, \\ \gamma &= -m\rho \dot{\rho}. \end{aligned}\right. \label{37c}
\end{equation}%
Thus, the invariant can be explicitly represented as: 
\begin{equation}
I=\frac{1}{2}\left[ \left( \frac{1}{\rho ^{2}}+m^{2}\dot{\rho}^{2}\right) r^{2}+\rho ^{2}\left( \mathbf{P}^{2}+\frac{\delta \left( \delta -1\right)-2\nu _{1}\nu _{2}\left( 1-R_{1}R_{2}\right) }{r^{2}}\right) -\rho \dot{\rho}m\left( r\mathcal{P}+\mathcal{P}r\right) \right] ,  \label{38}
\end{equation}%
where $\rho $ obeys the Ermakov-Pinney equation, which is given by: 
\begin{equation}
\ddot{\rho}+\frac{\dot{m}}{m}\dot{\rho}+\Omega ^{2}\rho =\frac{1}{m^{2}\rho
^{3}}.  \label{38a}
\end{equation}%
To solve the eigenvalue equation presented in Eq. \eqref{I1}, where the invariant is given in Eq. \eqref{38}, we introduce the following unitary transformation%
\begin{equation}
\mathcal{F}(r,\theta ,t)=U\left( r\right) \mathcal{G}(r,\theta ),  \label{40}
\end{equation}%
with the unitary operator
\begin{equation}
U\left( r\right) =\exp \left( \frac{im\dot{\rho}}{2\rho }r^{2}\right).
\label{41}
\end{equation}%
This transformation leads to the the operator $I$ being modified to $I^{\prime }$, according to the relation $I^{\prime }$:
\begin{equation}
I^{\prime }=U^{+}IU=\frac{1}{2}\left[ \rho ^{2}\left( \mathbf{P}^{2}+\frac{\delta
\left( \delta -1\right) -2\nu _{1}\nu _{2}\left( 1-R_{1}R_{2}\right) }{r^{2}}%
\right) +\frac{1}{\rho ^{2}}r^{2}\right] .  \label{42}
\end{equation}%
Consequently, Eq. \eqref{I1} maps to
\begin{equation}
I^{^{\prime }}(t)\mathcal{G}(r,\theta )=\varepsilon \mathcal{G}(r,\theta ).
\label{43}
\end{equation}%
Assuming
\begin{equation}
    \mathcal{G}(r,\theta )=r^{-\delta }\mathcal{Q}(r)\Theta _{\epsilon
}\left( \theta \right) ,
\end{equation}
the wave equation given in Eq. (\ref{43}) separates into two differential equations:
\begin{equation}
\mathcal{J}_{\theta }\Theta _{\epsilon }\left( \theta \right) =\lambda
_{\epsilon }\Theta _{\epsilon }\left( \theta \right) ,  \label{angular}
\end{equation}
\begin{equation}
\left[ \frac{\partial ^{2}}{\partial \varkappa ^{2}}-\varkappa ^{2}-\frac{\left(
\sigma _{l}^{\epsilon }\right) ^{2}-\frac{1}{4}}{\varkappa ^{2}}%
+2\varepsilon _{n,l}^{\epsilon }\right] \mathcal{Q}(\varkappa )=0,
\label{radial1}
\end{equation}%
in terms of the new variable $\varkappa =\frac{r}{\rho }$, and 
\begin{equation}
\sigma _{l}^{\epsilon }=\sqrt{\lambda _{\epsilon }^{2}+\left( \nu
_{1}+\epsilon \nu _{2}\right) ^{2}}.  \label{116}
\end{equation}
Here, $\epsilon =\epsilon _{1}\epsilon _{2}=\pm 1$, where each $\epsilon _{i}=\pm1,$ represents the eigenvalues associated with the reflection operators $R_{i}$. We then determine the eigenfunctions, $\Theta _{\epsilon }\left( \theta \right) $ and the corresponding eigenvalues $\lambda _{\epsilon }$ for two different cases:

\begin{itemize}

\item \textbf{First case}, $\epsilon =1$\textbf{:} 

This case corresponds to $\epsilon _{1}=\epsilon _{2}=1$ or $\epsilon _{1}=\epsilon _{2}=-1$ subcases, and the solution for $\Theta _{+1}\left( \theta \right) $ are given in \cite{Bouguerne20241} as follows: 
\begin{equation}
\Theta _{+1}\left( \theta \right) =a_{l}\mathbf{P}_{l}^{\left( \nu
_{1}+1/2,\nu _{2}+1/2\right) }\left( -2\cos \theta \right) \pm \acute{a}%
_{l}\sin \theta \cos \theta \mathbf{P}_{l-1}^{\left( \nu _{1}+1/2,\nu
_{2}+1/2\right) }\left( -2\cos \theta \right) ,  \label{24}
\end{equation}%
where $\mathbf{P}_{l}^{\left( \alpha ,\beta \right) }$ are Jacobi
polynomials and the eigenvalue is:
\begin{equation}
\lambda _{+}=\pm 2\sqrt{l\left( l+\nu _{1}+\nu _{2}\right) },\qquad
l\in  \mathbb{N}^{\ast }.  \label{25}
\end{equation}

\item \textbf{Second case, }$\epsilon =-1$\textbf{:} 

This case involves two sub-cases: $\left( \epsilon _{1},\epsilon _{2}\right) =\left( +1,-1\right) $
or $\left( \epsilon _{1},\epsilon _{2}\right) =\left( -1,+1\right) .$ Here,
the angular eigenfunction $\Theta _{-1}\left( \theta \right) $ is given by:%
\begin{equation}
\Theta _{-1}\left( \theta \right) =b_{l}\mathbf{P}_{l-1/2}^{\left( \nu
_{1}+1/2,\nu _{2}-1/2\right) }\left( -2\cos \theta \right) \pm \acute{b}%
_{l}\sin \theta \mathbf{P}_{l-1/2}^{\left( \nu _{1}-1/2,\nu _{2}+1/2\right)
}\left( -2\cos \theta \right) ,  \label{26}
\end{equation}%
and the corresponding eigenvalue is: \begin{equation}
\lambda _{-}=\pm 2\sqrt{\left( l+\nu _{1}\right) \left( l+\nu _{2}\right) }%
,\ \text{\ \ }l\in \left\{ 1/2,3/2,5/2, \cdots\right\} .  \label{27}
\end{equation}%
\end{itemize}
Subsequently, the solution of Eq. (\ref{radial1}) is given by:
\begin{equation}
\mathcal{Q}(\varkappa )=\mathcal{C}_{r}\varkappa ^{\sigma _{l}^{\epsilon }+%
\frac{1}{2}}e^{-\frac{\varkappa ^{2}}{2}}L_{n}^{\sigma _{l}^{\epsilon
}}\left( \varkappa ^{2}\right) ,  \label{28}
\end{equation}%
where the eigenvalues of the invariant are
\begin{equation}
\varepsilon _{n,l}^{\epsilon }=2n+\sigma _{l}^{\epsilon }+1.\text{ \ \ \ \ \ 
}\ n=0,1, \cdots .  \label{46}
\end{equation}
Here, $\mathcal{C}_{r}$ represents the normalization constant. 

\newpage

\section{Total quantum phase and wave function} \label{sec4}

By applying a unitary transformation and utilizing the Ermakov-Pinney equation, we simplify the expression in Eq. \eqref{PH}, yielding:
\begin{equation}
\dot{\eta}\left( t\right) =-\frac{\varepsilon _{n,l}}{m\rho ^{2}}+\Big{\langle} \mathcal{F}(r,\theta ,t)\Big{\vert} i\frac{\partial }{\partial
t}-\frac{\dot{\rho}}{2\rho }\left( r\mathcal{P}+\mathcal{P}r\right)
\Big{\vert} \mathcal{F}(r,\theta ,t))\Big{\rangle}.  \label{diff phase}
\end{equation}%
where it is straightforward to show that:
\begin{equation}
\Big{\langle} \mathcal{F}(r,\theta ,t)\Big{\vert} i\frac{\partial }{\partial t%
}-\frac{\dot{\rho}}{2\rho }\left( r\mathcal{P}+\mathcal{P}r\right)
\Big{\vert} \mathcal{F}(r,\theta ,t)\Big{\rangle} =0,  \label{102}
\end{equation}%
Thus, the phase can be written as
\begin{equation}
\eta _{l,n}^{\epsilon }\left( t\right) =-\left( 2n+\sigma _{l}^{\epsilon
}+1\right) \int^{t}\frac{dt^{\prime }}{m\left( t^{\prime }\right) \rho
\left( t^{\prime }\right) ^{2}},  \label{105}
\end{equation}%
Combining these results, we find that the total time-dependent wave function for Eq. \eqref{1} takes the form:
\begin{eqnarray}
\psi _{n,l,m_{s}}^{\epsilon _{1},\epsilon _{2}}\left( r,\theta ,t\right) &=&%
\mathcal{C}_{r}\frac{r^{\sigma _{l}^{\epsilon }-\delta +\frac{1}{2}}}{\rho
^{\sigma _{l}^{\epsilon }+\frac{1}{2}}}\exp \left\{ \left( im\rho \dot{\rho}%
-1\right) \frac{r^{2}}{2\rho ^{2}}\right.  \notag \\
&&\left. +i\left[ \frac{1}{2}\left( \lambda _{\epsilon }-m_{s}\left( 1+\nu
_{1}\epsilon _{1}+\nu _{2}\epsilon _{2}\right) \right) \int\limits^{t}\omega
_{c}\left( t^{\prime }\right) dt^{\prime }-\left( 2n+\sigma _{l}^{\epsilon
}+1\right) \int^{t}\frac{dt^{\prime }}{m\left( t^{\prime }\right) \rho
\left( t^{\prime }\right) ^{2}}\right] \right\}  \notag \\
&&\times L_{n}^{\sigma _{l}^{\epsilon }}\left( \frac{r^{2}}{\rho ^{2}}%
\right) \Theta _{\epsilon }\left( \theta \right) \chi _{m_{s}}.
\label{total wave function}
\end{eqnarray}%
Using the total wave function and Eq. (\ref{109}), we obtain the total phase factor $\mu _{l,n}^{\epsilon }\left(t\right) $in the following form:
\begin{equation}
\mu _{l,n}^{\epsilon }\left( t\right) =\frac{1}{2}\left( \lambda _{\epsilon
}-m_{s}\left( 1+\nu _{1}\epsilon _{1}+\nu _{2}\epsilon _{2}\right) \right)
\int\limits^{t}\omega _{c}\left( t^{\prime }\right) dt^{\prime }-\left(
2n+\sigma _{l}^{\epsilon }+1\right) \int^{t}\frac{dt^{\prime }}{m\left(
t^{\prime }\right) \rho \left( t^{\prime }\right) ^{2}}.  \label{phase T}
\end{equation}%

\noindent We also note that the wave functions in Eq. (\ref{total wave function}) depend on the values of $\epsilon$, so the solutions have to be distinguished by two cases:

\begin{itemize}

\item  \textbf{First case}, $\epsilon =1$\textbf{:}

In this case,  we have $\sigma _{l}^{+}=\sqrt{\lambda
_{+}^{2}+\left( \nu _{1}+\nu _{2}\right) ^{2}}$ and $\lambda _{+}=\pm 2\sqrt{%
l\left( l+\nu _{1}+\nu _{2}\right) }.$

\begin{enumerate}
    \item For $\epsilon _{1}=1,$ $\epsilon _{2}=1,$ we get
\small
\begin{eqnarray}
\psi _{n,l,m_{s}}^{+,+}\left( r,\theta ,t\right) &=&\mathcal{C}_{r}\frac{%
r^{\sigma _{l}^{+}-\delta +\frac{1}{2}}}{\rho ^{\sigma _{l}^{+}+\frac{1}{2}}}%
\exp \left\{ \left( im\rho \dot{\rho}-1\right) \frac{r^{2}}{2\rho ^{2}}%
\right.  \notag \\
&+i&\left. \left[ \frac{1}{2}\left( \lambda _{+}-m_{s}\left( 1+\nu _{1}+\nu
_{2}\right) \right) \int\limits^{t}\omega _{c}\left( t^{\prime }\right)
dt^{\prime }-\left( 2n+\sigma _{l}^{+}+1\right) \int^{t}\frac{dt^{\prime }}{%
m\left( t^{\prime }\right) \rho \left( t^{\prime }\right) ^{2}}\right]
\right\}  \notag \\
&\times& L_{n}^{\sigma _{l}^{+}}\left( \frac{r^{2}}{\rho ^{2}}\right) \Theta
_{+1}\left( \theta \right) \chi _{m_{s}}.  \label{wave1}
\end{eqnarray}

    \item For $\epsilon _{1}=-1,\epsilon _{2}=-1,$ we get
\small
\begin{eqnarray}
\psi _{n,l,m_{s}}^{-,-}\left( r,\theta ,t\right) &=&\mathcal{C}_{r}\frac{%
r^{\sigma _{l}^{+}-\delta +\frac{1}{2}}}{\rho ^{\sigma _{l}^{+}+\frac{1}{2}}}%
\exp \left\{ \left( iM\rho \dot{\rho}-1\right) \frac{r^{2}}{2\rho ^{2}}%
\right.  \notag \\
&+i&\left. \left[ \frac{1}{2}\left( \lambda _{+}-m_{s}\left( 1-\nu _{1}-\nu
_{2}\right) \right) \int\limits^{t}\omega _{c}\left( t^{\prime }\right)
dt^{\prime }-\left( 2n+\sigma _{l}^{+}+1\right) \int_{0}^{t}\frac{dt^{\prime
}}{M\left( t^{\prime }\right) \rho \left( t^{\prime }\right) ^{2}}\right]
\right\}  \notag \\
&\times& L_{n}^{\sigma _{l}^{+}}\left( \frac{r^{2}}{\rho ^{2}}\right) \Theta
_{+1}\left( \theta \right) \chi _{m_{s}}.  \label{wave2}
\end{eqnarray}
\end{enumerate}

\item  \textbf{Second case}, $\epsilon =-1$\textbf{:} 

In this case, we have  $\sigma _{l}^{-}=\sqrt{\lambda_{-}^{2}+\left( \nu _{1}-\nu _{2}\right) ^{2}}$ and $\lambda _{-}=\pm 2\sqrt{%
\left( l+\nu _{1}\right) \left( l+\nu _{2}\right) }.$

\begin{enumerate}
    
    \item For $\epsilon _{1}=-1,$ $\epsilon _{2}=1,$ we have
\small    
\begin{eqnarray}
\psi _{n,l,m_{s}}^{-,+}\left( r,\theta ,t\right) &=&\mathcal{C}_{r}\frac{%
r^{\sigma _{l}^{-}-\delta +\frac{1}{2}}}{\rho ^{\sigma _{l}^{-}+\frac{1}{2}}}%
\exp \left\{ \left( im\rho \dot{\rho}-1\right) \frac{r^{2}}{2\rho ^{2}}%
\right.  \notag \\
&+i&\left. \left[ \frac{1}{2}\left( \lambda _{-}-m_{s}\left( 1-\nu _{1}+\nu
_{2}\right) \right) \int\limits^{t}\omega _{c}\left( t^{\prime }\right)
dt^{\prime }-\left( 2n+\sigma _{l}^{-}+1\right) \int^{t}\frac{dt^{\prime }}{%
m\left( t^{\prime }\right) \rho \left( t^{\prime }\right) ^{2}}\right]
\right\}  \notag \\
&\times& L_{n}^{\sigma _{l}^{-}}\left( \frac{r^{2}}{\rho ^{2}}\right) \Theta
_{-1}\left( \theta \right) \chi _{m_{s}}.  \label{wave 3}
\end{eqnarray}

    \item For $\epsilon _{1}=1,$ $\epsilon _{2}=-1,$ we have

\small
\begin{eqnarray}
\psi _{n,l,m_{s}}^{+,-}\left( r,\theta ,t\right) &=&\mathcal{C}_{r}\frac{%
r^{\sigma _{l}^{-}-\delta +\frac{1}{2}}}{\rho ^{\sigma _{l}^{-}+\frac{1}{2}}}%
\exp \left\{ \left( iM\rho \dot{\rho}-1\right) \frac{r^{2}}{2\rho ^{2}}%
\right.  \notag \\
&+i&\left. \left[ \frac{1}{2}\left( \lambda _{-}-m_{s}\left( 1+\nu _{1}-\nu
_{2}\right) \right) \int\limits^{t}\omega _{c}\left( t^{\prime }\right)
dt^{\prime }-\left( 2n+\sigma _{l}^{-}+1\right) \int^{t}\frac{dt^{\prime }}{%
M\left( t^{\prime }\right) \rho \left( t^{\prime }\right) ^{2}}\right]
\right\}  \notag \\
&\times& L_{n}^{\sigma _{l}^{-}}\left( \frac{r^{2}}{\rho ^{2}}\right) \Theta
_{-1}\left( \theta \right) \chi _{m_{s}}.  \label{wave4}
\end{eqnarray}

\end{enumerate}

\end{itemize}

\normalsize
For the time-independent case, where $\omega _{c}=C^{st}$ and $m=C^{st}$, we can take  $\omega
\left( t\right) \rightarrow 0$. Thus, we get
\begin{equation}
\Omega =\frac{\omega _{c}}{2},\text{ \ }\rho =\frac{1}{\sqrt{\frac{1}{2}%
\omega _{c}m}},  \label{constants}
\end{equation}
and the total phase $\mu _{l,n}^{\epsilon }\left( t\right) $ becomes the same with the energy
spectrum $E_{n,l}^{\epsilon }$ of the stationary Dunkl-Pauli oscillator equation
\begin{equation}
\mu _{l,n}^{\epsilon }\left( t\right) =-E_{n,l}^{\epsilon }t,
\label{phase t}
\end{equation}
with
\begin{equation}
E_{n,l}^{\epsilon }=\frac{\omega _{c}}{2}\Big[ 2n+\sigma _{l}^{\epsilon}+1-\lambda _{\epsilon }+m_{s}\left( 1+\nu _{1}\epsilon _{1}+\nu
_{2}\epsilon _{2}\right) \Big] .  \label{energy}
\end{equation}

\noindent This energy spectrum result depends on the eigenvalues $\epsilon_1$ and $\epsilon_2$ of the reflection operators $R_1$ and $R_2$, which aligns with the findings presented in reference \cite{Bouguerne20241}.

\section{Conclusion}

In this study, we investigated the time-dependent Dunkl-Pauli oscillator in two dimensions, utilizing the Pauli equation as our foundational framework. By introducing the Dunkl formalism, we successfully constructed the Dunkl-Pauli equation in Cartesian coordinates, incorporating reflection symmetries and deformed momentum operators. This formulation enabled us to derive the Dunkl-Pauli Hamiltonian, influenced by a time-varying magnetic field and a harmonic oscillator characterized by time-dependent mass and frequency. Subsequently, we expressed the modified Hamiltonian in polar coordinates. Through an analysis of the eigenvalues and eigenfunctions of the Dunkl angular operator, we obtained exact wave function solutions using the Lewis-Riesenfeld invariant method. Additionally, we derived the total quantum phase factors for various reflection cases. Our results underscored the distinctive impact of Dunkl operators on the quantum system, providing precise expressions for the wave functions and energy eigenvalues. These findings illustrate the potential of Dunkl operators to deepen our understanding of quantum systems with deformed symmetries, paving the way for future research in quantum mechanics and mathematical physics.


\section*{Acknowledgments}

B. C. L. is grateful to Excellence Project PřF UHK 2211/2023-2024 for the financial support.

\section*{Data Availability Statement} 

Data sharing is not applicable to this article as no new data were created or analyzed in this study.

\section*{Conflict of Interests} 

The authors have no conflicts to disclose.

\end{document}